\documentclass[12pt,preprint]{aastex} 
 
\slugcomment{} 
 
\shorttitle{Time Variation of RM gradient} 
\shortauthors{Asada et al.} 
 
\begin{document} 
 
 
\title{Time Variation of Rotation Measure Gradient in 3C 273 Jet} 
 
 
\author{Keiichi Asada\altaffilmark{1, 2, 3}, Makoto Inoue\altaffilmark{1}, Seiji Kameno\altaffilmark{1,4} and Hiroshi Nagai\altaffilmark{1,2}} 
\affil{National Astronomical Observatory of Japan} 
\email{asada@vsop.isas.jaxa.jp} 
 
 
\altaffiltext{1}{National Astronomical Observatory of Japan} 
\altaffiltext{2}{Department of Astronomical Science, The Graduate University for Advanced Studies} 
\altaffiltext{3}{Institute of Space and Astronautical Science, Japan Aerospace Exploration Agency, \\  
3-1-1 Yoshinodai, Sagamihara, Kanagawa, 229-8510, Japan} 
\altaffiltext{4}{Department of Physics, Faculty of Science, Kagoshima University}

 
\begin{abstract} 
 
The existence of a gradient in the Faraday rotation measure (RM) of the quasar 3C 273 jet is confirmed by follow-up 
observations. A gradient transverse to the jet axis is seen for more than 20 mas in projected distance. Taking account 
of the viewing angle, we estimate it to be more than 100 pc. Comparing to the distribution of the RM in 1995, we detect 
a time variation of it at the same distance from the core over 7 yr. We discuss the origin of the Faraday rotation 
based on this rapid time variation. We rule out foreground media such as a narrow-line region, and suggest a helical 
magnetic field in the sheath region as the origin of this gradient of the  RM. 
 
\end{abstract} 
 
 
 

 
\section{Introduction} 
 
A gradient of the Faraday rotation measure (RM) across a jet 
is growing evidence for the existence of a toroidal or helical magnetic 
field associated with the jet. The first evidence for such a 
gradient of the RM across a jet was found by VLBA polarimetry 
toward the VLBI jet of a well-known quasar, 3C 273 (Asada 
et al. 2002, hereafter A02). Following this report, the same kind 
of gradient of the RM was reported for several jets of BL Lac 
objects (Gabuzda et al. 2004), and the gradient of the RM across 
the 3C 273 jet itself was also confirmed by several observations 
(Zavala \& Taylor 2005; Attridge et al. 2005). The role of a toroidal 
or helical magnetic field has been discussed for the launching 
and propagating mechanisms of jets based on magnetohydrodynamics 
from the theoretical point of view (e.g., Meier et al. 
2001 and references therein), and it has been suggested that the 
presence of a toroidal or helical magnetic field could be observed 
as a gradient of the RM across the jet (Blandford 1993). Recently, 
it has also been shown that the toroidal magnetic field in a jet's rest frame would be observed as a toroidal magnetic field 
in the observer frame with a compression of the pitch angle (Lyutikov et al. 2005). In this paper we report on our follow-up observation, which confirms our initial results and indicates a time variation.   
Throughout this paper, we use a Hubble constant of H$_{0}$ = 100 km s$^{-1}$ Mpc$^{-1}$ and a deceleration parameter of q$_{0}$ = 0.5 in order to keep consistency to the previous papers (e.g., A02).  An angular resolution of one milli-arcsecond (mas) corresponds to a linear resolution of 1.86 pc. 
 
\section{Observations and Data Reductions} 
 
Observations were carried out on 2002 December 15 using all 10 stations of the VLBA. 
Spacing the sampling by multiples 
of the fundamental separation in $\lambda^{2}$ space was useful in order to 
avoid the 2$\pi$ ambiguity when we measured the RM at discrete 
observing wavelengths. In order to arrange the observing wavelengths 
in $\lambda^{2}$ space within an observing band, we chose intermediate 
frequencies ( IFs) of 4.618, 4.688, 4.800, and 5.093 GHz 
in the 5 GHz band, and 8.118, 8.188, 8.402, and 8.593 GHz in the 
8 GHz band. Each IF had an 8 MHz bandwidth. Both left and 
right circular polarizations were recorded at each station. The integration 
time toward 3C 273 was 66 minutes at each frequency 
band. We observed OQ 208 as an instrumental calibration source, 
and 3C 279 and 4C 29.45 as polarization position angle calibration 
sources. 
 
An a priori amplitude calibration for each station was derived 
from a measurement of the antenna gain and system temperatures 
during each run. Fringe fitting was performed on each IF 
and polarization independently using the AIPS task fring. After 
deriving the delay and rate difference between parallel-hand cross-correlations 
(between LHCP-LHCP or RHCP-RHCP), the cross-hand 
correlations (between LHCP-RHCP) were fringe-fitted to 
determine the cross-hand delay difference. Once the cross-hand 
delay difference was determined, full self-calibration was performed 
for the parallel-hand cross-correlations. Images were initially 
obtained using DIFMAP, and then imported into AIPS to 
self-calibrate the full data sets using the task calib to get a final 
DIFMAP image. The instrumental polarizations of the antennas 
were determined for each IF at each band with OQ 208, using 
the AIPS task lpcal. The polarization angle offset at each station 
was calibrated using observations of 3C 279 obtained in the 
VLA/VLBA Polarization Calibraton Monitoring Program (Myers 
 \& Taylor\footnote{See http://www.vla.nrao.edu/astro/calib/polar/}). 
The source was observed on 2002 December 17 by 
the VLA at 4.8851 and 4.8451 GHz, and 8.4351 and 8.4851 GHz. 
Any change in the source between the two observations is presumably 
small, since the observations were made with the VLA 
and VLBA with in 2 days of each other. The observing frequencies 
were slightly different between the VLA and VLBA observations, 
and the VLBA polarization position angles were interpolated using 
the VLA polarization position angle.  
In order to obtain the distributions of RM and projected magnetic field, we restored images at higher frequencies to match the resolution at the lowest frequency observation. The restored beam size was 3.22 mas $\times$ 1.23 mas with the major axis at a position angle of - 4.$^{\circ}$63.  
 
For the registration of images at different frequencies, we identified 
four distinct components. We measured the relative position 
with respect to the core component, peak flux, integrated flux, and 
size with the AIPS task imfit. As the core position may shift with 
frequency because of synchrotron self-absorption (Blandford \& 
K\"onigl 1979), we used optically thin components to register the 
images at different frequencies. We derived the positions of these 
components by weighted signal-to-noise ratios. The variances of 
the positions of the optically thin components are 0.059 mas in 
right ascension and 0.061 mas in declination. Those are 0.05 and 
0.02 times the beam size, respectively. The distribution of the 
 RM  was obtained by the AIPS task RM with polarization images 
at 4.618, 5.093, 8.118, and 8.593 GHz, with regions where the 
polarized intensity is greater than 3 times the  rms noise in the 
polarized intensity. 
 
\section{Results} 
 
\subsection{Apparent Motions} 
 
The distribution of the RM is shown superposed on the distribution 
of the total intensity at the first (A02) and second epochs 
in Figure 1. Five components in the jet are identified, and are 
labeled in the same manner as at the first epoch.   
Apparent velocities $\beta_{app}$ are measured with respect to the core component D and the detailed parameters are listed  in table 1.  
The components C1, 
C2, and C3 correspond to the components F, D, and B in independent 
measurements from the NRAO 2 cm survey (Kellermann 
et al. 2004), respectively. The measured apparent velocities are 
in good agreement with each other. The velocities measured by 
Kellermann et al. (2004) are used in the following discussion, 
since the velocities measured by ourselves are based on just two epoch 
measurements at lower frequency.

The viewing angle of the jet can be constrained from the Doppler effect by equation  $\theta_{max} = 2 \arctan{1/\beta_{app}}$, where $\theta_{max}$ is the upper limit of the viewing angle between the jet and line of sight. The $\theta_{max}$ for components C1, C2 and C3 is 23.5$^{\circ}$ $\pm$ 2.1$^{\circ}$, 22.6$^{\circ}$ $\pm$ 2.6$^{\circ}$, and 17.8$^ {\circ}$ $\pm$ 0.3$^{\circ}$, respectively. 
 
\subsection{ RM  distribution} 
 
The longer integration time toward 3C 273 brings us a better 
$u-v$ coverage at this epoch compared to the first epoch; thus we 
can reveal the distribution of the RM on a large part of the jet. We 
show the cross section of the RM at the second epoch along several 
lines perpendicular to the jet in Figure 2, and the RM distribution 
in a bird's-eye view in Figure 3. We detected gradients 
in the RM up to 20 mas from the core at the second epoch, which 
corresponds to 38.4 pc in projected distance. Taking into account 
the upper limit of the viewing angle of the jet of component C2 
of 22.6, the linear distance is longer than 100 pc. 
 
The value of the RM is typically a few hundred rad m$^{-2}$, 
always positive, and the RM on the left side is larger by a few 
hundred rad m$^{-2}$ than on the right side. The trend of the gradient 
of the RM is consistent with our previous results (A02) and independent 
results by Zavala \& Taylor (2005) and Attridge et al. 
(2005). However, the typical value of the RM observed by Zavala 
\& Taylor is in the range from 0 to 2000 rad m$^{-2}$ in the region 
where we detect the RM gradient. This RM value is obviously 
larger than given by our observations. Two possibilities are suggested. 
One is that there is an averaging effect for low-frequency 
observations (Zavala \& Taylor 2005). The other is that there is a 
real time variation of the distribution of the RM (Attridge et al. 
2005). The possibility of a time variation of the RM is discussed 
in the following section. 
 
On the other hand, even in our RM map only, a decreases in the RM from the core to the jet  is seen (see Fig 3). 
This tendency is also consistent with RM of 22000 rad m$^{-2}$ at innermost components of 1 mas from the Stokes $I$ peak of the map  
by Attridge et al. (2005). 
 
 
\section{Discussions} 
 
Since the amount of Faraday rotation is larger than 90$^{\circ}$  and the 
fractional polarization is reasonably strong, the magnetized plasma 
which is responsible for this Faraday rotation should be located 
in front of the emitting region. What makes this RM gradient? 
Two possibilities can be addressed. One is that the plasma is 
closely associated with the visible jet, as suggested by A02. The 
other is that foreground plasma independent of the jet produces 
the RM gradient by chance, as suggested by G\'omez et al. (2000). 
Time variation of the RM distribution could address this issue. 
 
First we consider the case in which the RM is produced by a 
magnetized plasma in a foreground screen that has nothing to do 
with the jet itself. Even a magnetized plasma in a narrow-line 
region (NLR) may not show variations in the Faraday screen 
on these short timescales, for the following reasons. It is well 
known that magnetized plasma in a NLR typically has a velocity 
of 1000 km $s^{-1}$. 
At the distance of 3C 273, this motion corresponds 
to 0.00056 mas yr$^{-1}$.  
The estimated motion between the 
two epochs (7 yr apart) would be 0.004 mas. This is too small to 
detect, and would not cause time variation in a foreground screen. 
Also, a change of the density in the magnetized plasma and/or 
magnetic field could not produce a time variation. Therefore, the 
characteristic timescale of the change in the RM should be related 
to the size of the magnetized plasma and the sound velocity 
or Alfv\'en velocity.  
Assuming a plasma temperature in the NLR 
of 10$^{4}$ K and an equipartition condition for the magnetic pressure 
and thermal pressure, we estimate the Alfv\'en velocity and the sound speed to be 4 $\times$ 10$^{-5}$ c and 7 $\times$ 10$^{-5}$ c, respectively.  
If we assume a scale length for the magnetized plasma of 1 pc, the typical 
timescale of the variation in the foreground screen is estimated 
to be 8 $\times$ 10$^{4}$ yr and 4 $\times$ 10$^{4}$ yr, respectively.   
 
Second we consider the RM being caused by magnetized 
plasma associated with the jet itself. In this case we expect the 
sheath around the emitting jet to be the origin of the RM (Inoue 
et al. 2003), since we do not see internal Faraday rotation. The 
sheath would be slower than the spine, but would have a relativistic 
velocity, since we do not detect any emission from the sheath 
at the counterjet. In addition, a highly time variable RM distribution 
would be expected due to the interaction between the spine 
and the sheath (Wardle et al. 2006). 
 
We show the cross section of the RM at a distance of 9 mas 
from the core at both epochs in Figure 4 (see also Fig. 1). The 
slice of RM in 2002 is clearly different from that in 1995, and we 
think we detect time variation of the RM toward the jet between 
these epochs. In addition, we evaluate the registration error effect 
on the time variation of the RM as follows. First we assume 
that there are no time variations of the RM distribution between 
the two epochs, but there may be apparent time variations due 
to the registration error. In this case we would find a position 
where the RM distributions show a good pattern matching within 
the registration error. For this purpose, we tentatively shift the 
distribution of the RM of the first epoch up to $\pm$ 1 mas in right ascension 
and declination with respect to that of the second epoch, 
and calculate the reduced $\chi^{2}$ for each shifted position. We note 
that 1 mas is large compared to the registration error of one-fifth 
of the synthesized beam width. 
This calculation was performed only for shifts of the pixels where both distributions of the 
RM had a significant value (SNR $>$ 3) and the integrated area was larger than 3 times the synthesized beam size of the first epoch.
The reduced $\chi^{2}$  larger than 4.0 was obtained all over the jet, suggesting 
that the two RM distributions are not identical, or that 
they changed with the jet evolution. 
 
 Zavala \& Taylor (2001, 2005) reported that they did not detect 
time variation in the front of the jet, while they detected it 
toward the core over 6 months. The difference between our result 
and theirs would simply be due to the difference of the span between 
the two epochs and the choice of the observation frequencies. 
Since we observed at lower frequencies with the VLBA, it 
made it easy to detect small differences in the  RM. In order to detect 
small differences in the  RM, observations at lower frequencies 
with high angular resolution are necessary. 
Even if the RM is produced by a magnetized plasma in a foreground 
screen, the RM may also change if a relativistically moving 
polarized jet component is seen through a patchy foreground 
Faraday screen (Zavala \& Taylor 2001). If this is the case, a 
repeatable distribution of the RM should be expected when the 
moving polarized jet component is located at the same position, 
as reported toward 3C 120 (G\'omez et al. 2000; J. L. G\'omez et al. 
2006, private communication). 
We cannot check for any repeatability of the Faraday screen 
effect, since our analysis is based on only two-epoch observations. 
Therefore, it is not possible to discriminate between these 
two possibilities; a helical magnetic field associated with the jet 
or enhancement of the density in a foreground plasma. If the  
deconvolved size of the emitting knot and the typical scale of the 
foreground screen were smaller than the beam size, a rapid time 
variation of the RM would be expected. As the deconvolved size 
of the emitting knot is larger than the beam size, we expect the 
emission from the moving polarized jet to be traveling through 
the same foreground plasma at both epochs. We think that the 
time variation of the RM gradient is not associated with an enhancement 
of the density in a foreground plasma; therefore, a 
helical magnetic field in the sheath is preferable as the origin 
of the RM gradient. As is discussed with the first-epoch result 
(A02), the RM gradient can be explained using a helical magnetic 
field. Since the RM at the southeast side of the jet is larger 
than that at the northwest side, the direction of the toroidal component 
is clockwise seen from the core toward the downstream 
of the jet. In addition, if we assume that the offset of the RM is 
ascribed to the longitudinal component of the helical magnetic 
field, the helicity of the field can be estimated to be right-handed. 
Thus, we could define the helicity of the helical magnetic field 
and the direction of the rotation of the accretion disk or spinning 
black hole itself as clockwise as we see it. Therefore, we propose 
that a helical magnetic field in the sheath is responsible for the 
time-variable  RM, and our monitoring program will answer this 
question, describing the characteristics of the time variation of 
the RM distribution in the 3C 273 jet. 
 
\section{Conclusions} 
 
In order to confirm the RM gradient across the 3C 273 jet, we 
made a follow-up observation using multifrequency VLBA polarimetry. 
The systematic gradient across the jet is confirmed for 
more than 100 pc along the jet, and the trend of the RM gradient 
is consistent with that revealed by previous observations. Since 
the amounts of the Faraday rotation exceed 90$^{\circ}$, the origin of the 
Faraday rotation should be in the foreground of the emitting jet. 
On the other hand, we detected a time variation in the distribution 
of the RM in comparison to that in 1995, and this rapid time variation 
rules out the possibility that a foreground magnetized cloud 
independent of the jet, such as a narrow-line region, is responsible 
for the origin of the Faraday rotation. Therefore, the sheath around 
the ultra-relativistic jet is likely to be the origin. 
 
\acknowledgments 
 
This research has made use of data taken by Very Long Baseline Array (VLBA). VLBA is operated by the National Radio Astronomy Observatory (NRAO), and NRAO is a facility of the National Science Foundation operated under cooperative agreement by Associated Universities, Inc.

\clearpage

\begin{figure}[htbp] 
\epsscale{1.0} 
\plotone{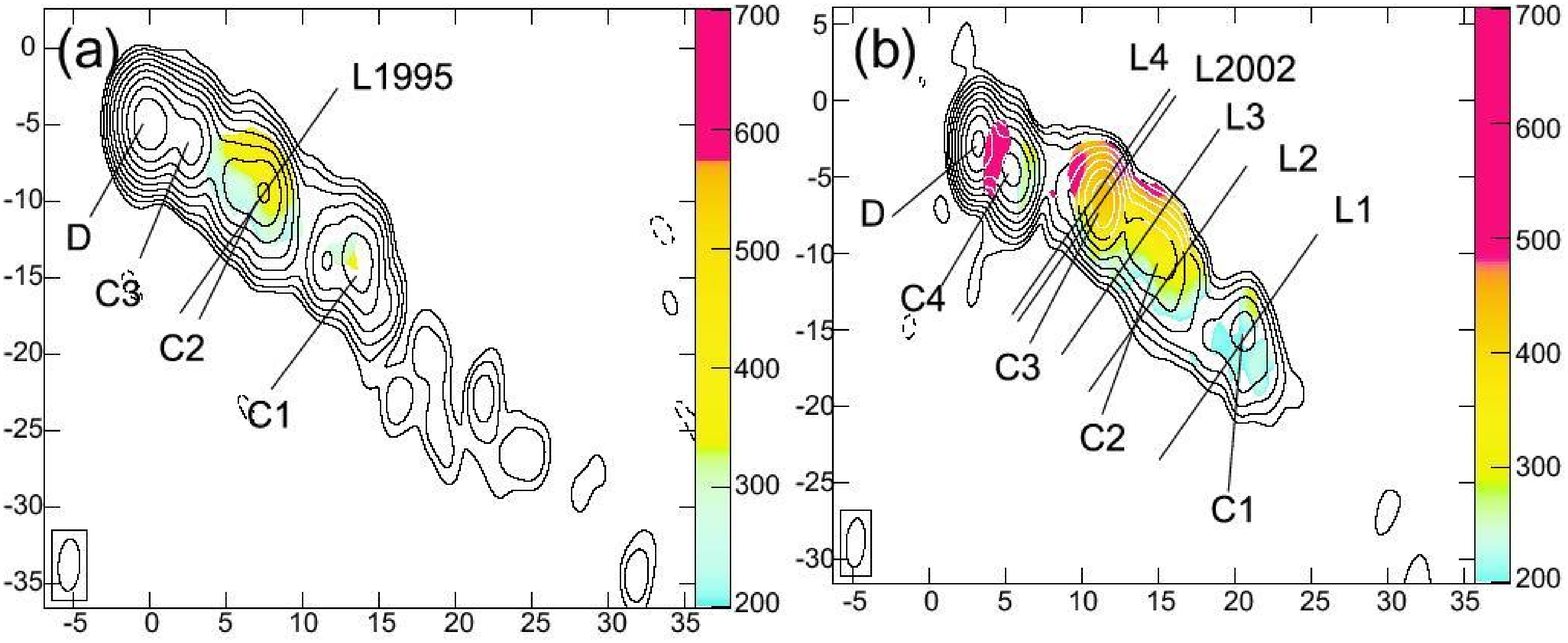} 
\caption{(a) Distribution of RM (color scale) superposed on the contour images of the total intensity at 4.760 GHz at the first epoch.   
Contours are plotted at -1, 1, 2, 4, 8, 16, 32, 64, 128, 256, 512 and 1024 $\times$ three-times  
the r.m.s. noise of that of the total intensity at 4.760  GHz. The synthesized beam size is 3.4 mas $\times$ 1.4 mas with  
the major axis at a position angle of - 3$^{\circ}$.7.  
(b) Distribution of RM (color scale) superposed on the contour images of the total intensity at 4.618 GHz at the second epoch.   
Contours are plotted at -1, 1, 2, 4, 8, 16, 32, 64, 128, 256, 512 and 1024 $\times$ three-times  
the r.m.s. noise of that of the total intensity at 4.618 GHz. The synthesized beam size is 3.2 mas $\times$ 1.2 mas with the  
major axis at a position angle of - 4$^{\circ}$.7.  
The  RMs are plotted in the region where the polarized intensity is greater than 3-times the r.m.s.  
noise in the polarized intensity. Line L1995 in figure 1 (a) and L1, L2, L3, L4, and L2002 in figure 1 (b) show the position of  
the cross-sections shown in figure 2 and figure 4. \label{figure:  RM }} 
\end{figure} 
 
\begin{figure}[htbp] 
\epsscale{1.0} 
\plotone{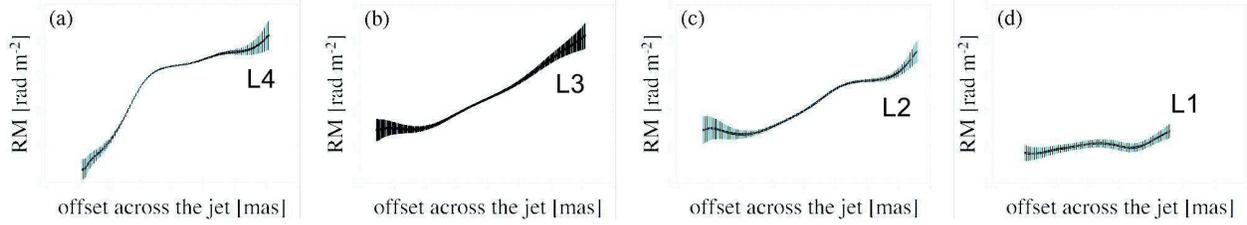} 
\caption{Cross sections of the RM distribution across (a) C3, (b) upper side of C2, (c)  lower side of C2, and (d) C1 in the second  
epoch derived using the AIPS task SLICE (see line L1, L2, L3 and L4 in figure 1). The shaded area along the curved line of  
 RM  indicates the standard deviation (1 $\sigma$) in  RM. The profile of the RM distribution is anti-symmetric with respect  
to the central axis of the jet.\label{fig1}} 
\end{figure} 
 
\begin{figure}[htbp] 
\epsscale{1.0} 
\plotone{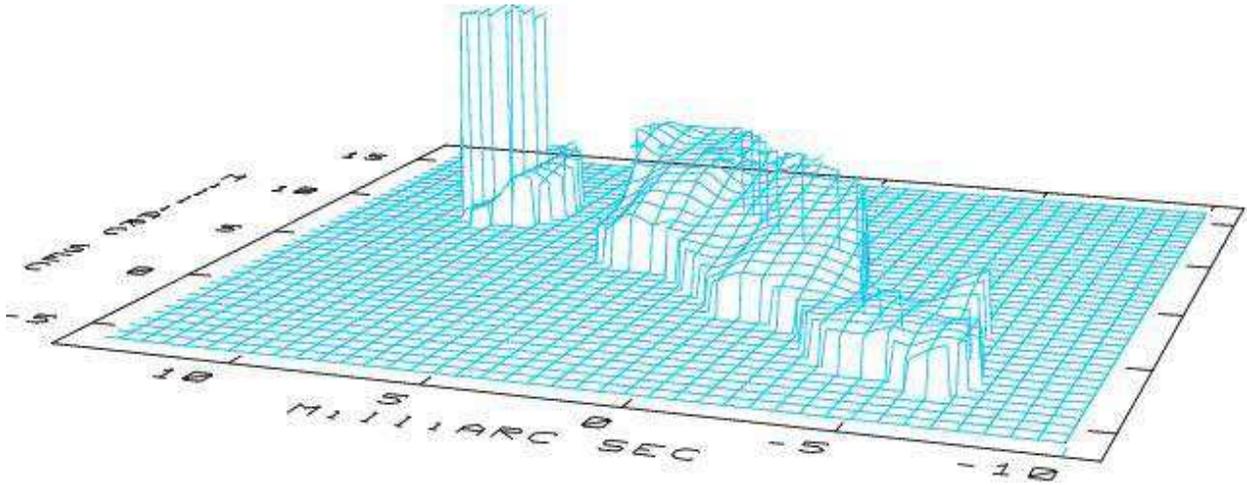} 
\caption{\label{figure: be RM }Bird's eye view of the distribution of  RM.  
Gradient of RM can be clearly seen in both transverse and along the jet axis.} 
\end{figure} 
 
\begin{figure}[htbp] 
\epsscale{1.0} 
\plotone{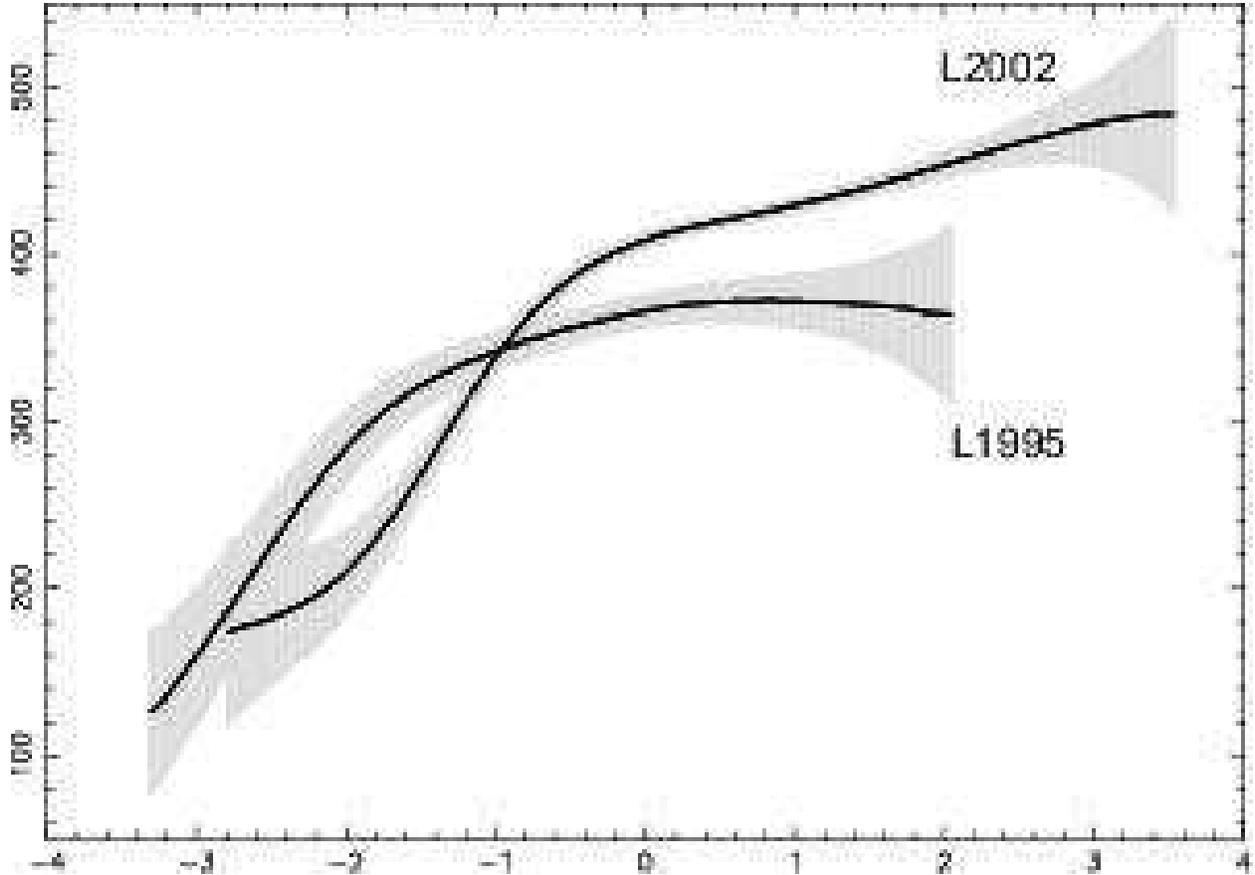} 
\caption{\label{figure: super}The cross sections of RM across the position at 9 mas from the core at both epochs are show (see line L1995 and L2002 in figure 1.).  
L1995 indicates the cross section at the first epoch and L2002 indicates that at the second epoch.   
The shade area shows three times the standard deviation (3 $\sigma$) in  RM. There are the same trend of the gradient across the jet.  
However, RM at the second epoch is obviously larger than that at the first epoch.} 
\end{figure}

\begin{table}[t]
\begin{center}
\caption{Apparent proper motion}
\begin{tabular}{cccc}
\hline

Component & $\beta_{app}$ & $\beta_{app}$ by KK04 & $\theta_{max}$ \\ 
\hline 
\hline

C1 & 5.9 $\pm$ 0.6 & 4.8 $\pm$ 0.9 & 23$^{\circ}$.5 $\pm$ 2$^{\circ}$.1 \\
C2 & 5.3 $\pm$ 0.6 & 5.0 $\pm$ 1.2 & 22$^{\circ}$.6 $\pm$ 2$^{\circ}$.6 \\
C3 & 7.2 $\pm$ 0.6 & 6.4 $\pm$ 0.2 & 17$^{\circ}$.8 $\pm$ 0$^{\circ}$.3 \\
\hline
\end{tabular}
\label{table:Apparent motion and Constraints for the viewing angle}
\end{center}
\end{table}

 
\end{document}